# The PEPSI Exoplanet Transit Survey (PETS). V: New Na D transmission spectra indicate a quieter atmosphere on HD 189733b

E. Keles,[1,2] S. Czesla,[3] K. Poppenhaeger,[1,4] P. Hauschildt,[5] T. A. Carroll,[1] I. Ilyin,[1] M. Baratella,[1,6] M. Steffen,[1] K. G. Strassmeier,[1,4] A. S. Bonomo,[7] B. S. Gaudi,[8] T. Henning,[9] M. C. Johnson,[8] K. Molaverdikhani,[10] V. Nascimbeni,[11] J. Patience,[12] A. Reiners,[13] G. Scandariato,[14] E. Schlawin,[15] E. Shkolnik,[12] D. Sicilia,[14] A. Sozzetti,[7] M. Mallonn,[1] C. Veillet,[16] J. Wang,[8] F. Yan[17]

[1] *Leibniz-Institut für Astrophysik Potsdam (AIP), An der Sternwarte 16, 14482 Potsdam, Germany*
[2] *Freie Universität Berlin, Institute of Geological Sciences, Malteserstr. 74-100, 12249 Berlin, Germany*
[3] *Thueringer Landessternwarte Tautenburg, Sternwarte 5, 07778 Tautenburg, Germany*
[4] *Institut für Physik & Astronomy, University of Potsdam, Karl-Liebknecht-Str. 24/25, 14476 Potsdam, Germany*
[5] *Hamburger Sternwarte, Gojenbergsweg 112, 21029 Hamburg, Germany*
[6] *European Southern Observatory, Alonso de Cordova 3107 Vitacura, Santiago de Chile, Chile*
[7] *INAF - Osservatorio Astrofisico di Torino, Via Osservatorio 20, 10025 Pino Torinese, Italy*
[8] *Department of Astronomy, The Ohio State University, 4055 McPherson Laboratory, 140 West 18th Ave., Columbus, OH 43210 USA*
[9] *Max-Planck-Institute for Astronomy, Königstuhl 17, 69117 Heidelberg, Germany*
[10] *Ludwig-Maximilians-Universität München (LMU), Geschwister-Scholl-Platz 1, 80539 München, Germany*
[11] *INAF – Osservatorio Astronomico di Padova, Vicolo dell'Osservatorio 5, 35122 Padova, Italy*
[12] *School of Earth and Space Exploration, Arizona State University, 660 S. Mill Ave., Tempe, Arizona 85281, USA*
[13] *Georg-August-Universitaet Göttingen, Friedrich-Hund-Platz 1, 37073 Göttingen, Germany*
[14] *INAF - Osservatorio Astrofisico di Catania, via S. Sofia 78, 95123 Catania, Italy*
[15] *Steward Observatory, University of Arizona, 933 N. Cherry Ave., Tucson, AZ 85721, USA*
[16] *Large Binocular Telescope Observatory, 933 N. Cherry Ave., Tucson, AZ 85721, USA*
[17] *Department of Astronomy, University of Science and Technology of China, Hefei 230026, China*

**ABSTRACT**

Absorption lines from exoplanet atmospheres observed in transmission allow us to study atmospheric characteristics such as winds. We present a new high-resolution transit time-series of HD 189733b, acquired with the PEPSI instrument at the LBT and analyze the transmission spectrum around the Na D lines. We model the spectral signature of the RM-CLV-effect using synthetic PHOENIX spectra based on spherical LTE atmospheric models. We find a Na D absorption signature between the second and third contact but not during the ingress and egress phases, which casts doubt on the planetary origin of the signal. Presupposing a planetary origin of the signal, the results suggest a weak day-to-nightside streaming wind in the order of 0.7 km/s and a moderate super-rotational streaming wind in the order of 3 - 4 km/s , challenging claims of prevailing strong winds on HD 189733b.

## 1 INTRODUCTION

High-resolution exoplanet transmission spectroscopy allows the deduction of atmospheric properties from absorption line shape properties such as the line contrast (LC), the line shift ($v_{shift}$), and the full width at half maximum (FWHM). The LC can be used for instance to probe exospheric properties (Huang et al. 2017; Gebek & Oza 2020) and escape (Yan & Henning 2018). The line shifts and line widths are used often to study atmospheric winds like day-to-nightside (dn) winds (Snellen et al. 2008) and super-rotational (sr) winds (Louden & Wheatley 2015) as well as vertical upward winds (Seidel et al. 2020). The dn winds introduce a blueshift to the absorption lines caused by the pressure-and-temperature gradi-ent between the planetary day-and-nightside showing wind speeds of ~ 1-10 km/s in the mbar regime (Knutson et al. 2008; Miller-Ricci Kempton & Rauscher 2012). The sr winds or vertical upward winds impose excess broadening on the absorption lines on top of the natural line broadening mechanisms as well as the planetary rotation (Seidel et al. 2020; Keles 2021). The former can show wind velocities of a few km/s at the 1bar regime (Showman et al. 2012), whereas the wind properties depend strongly on the atmospheric properties, energy budget, and drag within the atmosphere (Miller-Ricci Kempton & Rauscher 2012; Wardenier et al. 2021).

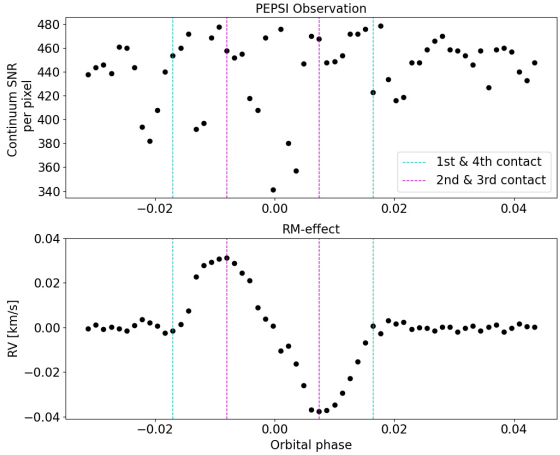

**Figure 1.** The HD 189733b transit observation. Top: Continuum signal-to-noise ratio per combined pixel (95% quantile). Bottom: The RM-effect. Dashed vertical lines show the contact points 1st-4th (cyan) and 2nd-3rd (purple).

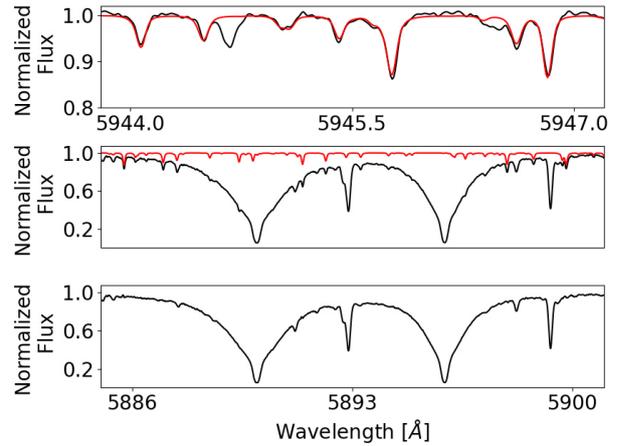

**Figure 2.** Example for the telluric correction of a PEPSI spectrum (black) using a template spectrum (red). The top panel shows the wavelength region used to determine the best-fit telluric template spectrum. The central panel shows the telluric template and the observed spectrum. The bottom panel shows the corrected spectrum.

Species with strong atmospheric fingerprints like sodium, which is the first detected species from a ground-based facility (Redfield et al. 2008) on an exoplanet, namely HD 189733b, are crucial to study winds. The atmospheric Na D- absorption observed during the transit of HD 189733b has been revisited by several studies (e.g. Wyttenbach et al. 2015; Louden & Wheatley 2015; Yan et al. 2017; Casasayas-Barris et al. 2017; Pino et al. 2018; Borsa & Zannoni 2018; Seidel et al. 2020; Langeveld et al. 2021; Sicilia et al. 2022; Langeveld et al. 2022) from the same HARPS data sets (for further information on the HARPS data see Wyttenbach et al. 2015). Comparing the inferred Na D-absorption line shape properties, the derived FWHM deviates by about 0.2 Å , and the LC deviates by about 0.15 % in the different studies. Such a deviation in FWHM results in very different wind strength estimations, and in particular HD 189733b is thought to show one of the strongest sr winds (Keles et al. 2020) or vertical upward wind (Seidel et al. 2020) or both for any exoplanet, deduced using the Na D-absorption lines. In the case of dn winds, although blueshifts of the Na D-absorption lines in the order of 2 km/s have been presented from the combined HARPS data (Louden & Wheatley 2015; Casasayas-Barris et al. 2017; Langeveld et al. 2022; Sicilia et al. 2022), the individual HARPS transit data sets show phase dependent unexpected large blue- and redshifts of the Na D-absorption lines (Louden & Wheatley 2015; Borsa & Zannoni 2018; Sicilia et al. 2022).

The uncertainties regarding the origin and amplitude of the different kinds of winds inferred from the Na D-lines on HD 189733b are calling for a study with a higher signal-to-noise (S/N) ratio dataset.

## 2 OBSERVATION, DATA REDUCTION AND TELLURIC CORRECTION

One transit of HD 189733b was observed at the LBT with the PEPSI spectrograph at a spectral resolution of 130 000 within the course of the PETS survey (Keles et al. 2022; Johnson et al. 2023; Scandariato et al. 2023; Petz et al. 2024). The observation was carried out on 2021-11-11 from 02:39 UT to 06:40 UT using an exposure time of 200 seconds, gathering 59 spectra, with 32 spectra observed out-of-transit and 27 spectra observed during the transit. The observation covered the wavelength ranges from around 4220 Å - 4770 Å and 5360 Å - 6314 Å in two different cross-dispersers. The former wavelength range is not used for the analysis. Figure 1 shows (top) the continuum S/N ratio which is around 450 per binned pixel (95% quantile) for the latter wavelength range which covers the Na D lines. Further information regarding the PEPSI spectrograph is provided in Strassmeier et al. (2015). The data reduction is applied as described in Keles et al. (2022). All spectra are resampled into an equal step size 0.01Å in wavelength scale and aligned at the stellar rest frame (with the stellar Na D line cores centered at 5889.950Å and 5895.924Å). The barycentric, as well as the stellar Keplerian motion, are corrected, leaving only the RM-effect induced (pseudo-) radial velocity shift, as shown in Figure 1 (bottom).

We correct the spectra for telluric water lines using *THOR* (Keles et al. 2019), which uses HAPI (Kochanov et al. 2016) to calculate absorption spectra using absorption coefficients provided by the HITRAN database (Gordon et al. 2017). THOR iterates HAPI for different optical path lengths and determines the best matching artificial water absorption spectrum (i.e. the telluric water template spectrum) by minimizing the $\chi^2$-value between each stellar spectrum and the telluric water template spectrum at the wavelength region 5943.00 Å to 5947.20 Å which is mostly free of stellar lines, as demonstrated for one PEPSI spectrum in Figure 2 (top panel). The parameters pressure, temperature and resolution are set to constant values that match best the observed telluric lines. After determining the best-fit optical path length, we calculate the telluric water template spectrum for the wavelength regions around the Na D lines (central panel, red line) and divide the observed spectrum by the telluric water template spectrum (bottom panel).

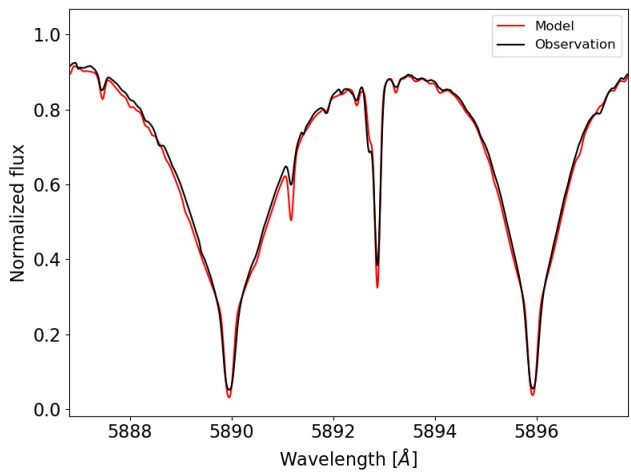

**Figure 3.** Comparison of the observed (black) and synthetic (red) Na D-lines in the mean out-of-transit spectrum.

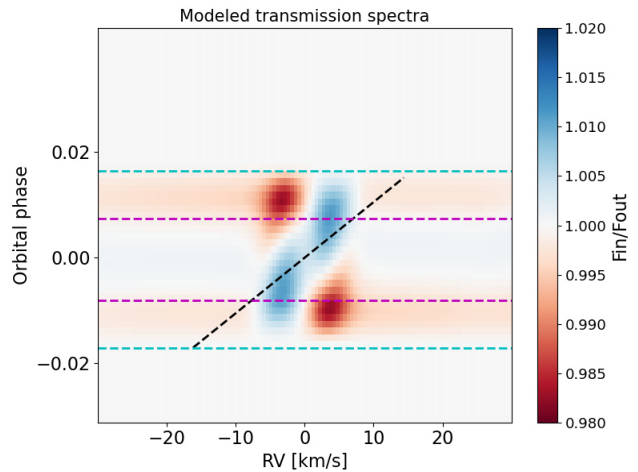

**Figure 4.** The modeled transmission spectra show the RM-CLV-feature at the Na D2-line for the different in-transit spectra in the stellar restframe. Cyan dashed lines mark the 1st - 4th contact and the purple dashed lines mark the 2nd - 3rd contact. The black dashed line illustrates the trajectory of planetary Na D2-absorption.

## 3 THE RM- AND CLV- MODEL

The RM- and CLV-effect introduce residuals into the observations and need to be considered. For this, specific intensities are derived using PHOENIX models (Hauschildt & Baron 1999) based on LTE stellar atmospheres. PHOENIX provides synthetic stellar specific intensity spectra for limb angles with $\mu = \cos\theta$ between $\mu = 0$ (limb) and $\mu = 1$ (center) from a grid of 127 angles. The spectra are broadened according to the instrumental response and spectral resolution. The PHOENIX model and spectra are from the NewEra model grid (Hauschildt et al, in prep.) which includes massive updates compared to the Husser et al. (2013) model grid, in particular for the line databases. The model and the radiation fields are calculated with spherical geometry.

We use planetary and stellar parameters from Bonomo et al. (2017) for the modelling, i.e. stellar radius $R_\star = 0.756 \pm 0.018\ R_\odot$, planetary radius $R_{pl} = 1.138 \pm 0.027\ R_{Jup}$, inclination inc = 85.58 $\pm$ 0.06°, orbital period P = 2.218577 $\pm$ 0.000010 days, semi-major axis a = 0.0310 $\pm$ 0.0006 AU and transit midpoint $T_{BJD-TDB}$ = 2453955.525551 $\pm$ 0.000009 days. We derive a planet velocity semi-amplitude (Kp) value of Kp = 152 km/s assuming a circular orbit. Recently, Cristo et al. (2023) investigated the transit of HD 189733b with ESPRESSO data acquired at the VLT, inferring differential stellar rotation with $v \sin i$ = 3.38 $\pm$ 0.06 km/s at the equator and $v \sin i$ = 2.58 $\pm$0.38 at the poles. However, we adapt the provided stellar rotation velocity $v \sin i$ = 3.5 $\pm$ 1.0 km/s within its uncertainty range to $v \sin i$ = 2.8 km/s which results in a better match of the modeled and observed RM-CLV- signatures in the combined transmission spectrum (see Section 3.1). We set the obliquity $\lambda = 0°$ as the planet has a nearly perpendicular orbit to the stellar rotation axis (Cameron et al. 2010).

The stellar disc is modeled with 100 x 100 pixels containing absolute specific intensities shifted in each pixel according to the stellar $v \sin i$ assuming rigid surface rotation. The $T_{eff}$ of HD 189733 is derived by different works ranging from e.g. $T_{eff}$ = 4875K $\pm$ 45K (Boyajian et al. 2015) to $T_{eff}$ = 5109K $\pm$ 146K (Mortier et al. 2013), with presumably solar abundance. In order to reproduce the observed Na D- lines, we chose $T_{eff}$ = 5000K and abundance value of A(Na) = 6.5 dex. Figure 3 shows the comparison of the observed and modeled stellar spectrum around the Na D-lines.

We compute the stellar spectrum at different orbital phase positions of the planet by summing over the stellar pixels during the planetary transit. We calculate the transmission spectra for the model by dividing all spectra by the combined out-of-transit transit spectrum. Figure 4 illustrates the modeled RM-CLV-feature in a 2D map around the Na D2-line. As the modeled transmission spectrum does not include atmospheric absorption from the planet, only the RM-CLV- feature becomes visible in the model. The dashed black line shows where we would expect the planetary absorption. The major part of the absorption overlaps with the RM-CLV- feature between the 2nd-3rd contact points. Between the ingress and egress i.e. the 1st-2nd contact point as well as the 3rd-4th contact point, the impact of the RM-CLV-effect at the expected position of the planet is small.

We aim to verify the match of the RM-CLV-model with the data. For instance, Casasayas-Barris et al. (2022) and Dethier & Bourrier (2023) showed for Mascara-1b that the stellar and planetary properties can be derived by comparing the observed and modeled RM-CLV-profile at the Na D- lines. As HD 189733b is expected to show significant Na D- absorption, we use a different approach to verify the RM-CLV-model. We first create the combined transmission spectrum by shifting the spectra to the planetary restframe and combining them. From this, we derive the mean line profile (Semel et al. 2009) by combining several lines, aiming to boost the S/N ratio, in the velocity space. To that end, the wavelengths are put into the same velocity grid ($\pm$50 km/s in steps of 0.1 km/s). We use 262 stellar absorption lines with a line depth of at least 0.4, excluding the Na D-lines as those might include the planetary absorption. Although the mean line profile cannot account for the modeling uncertainty introduced by the significantly different CLV for lines such as the Na D- lines, it enables to proof the average match of the RM-CLV-model with the observations.

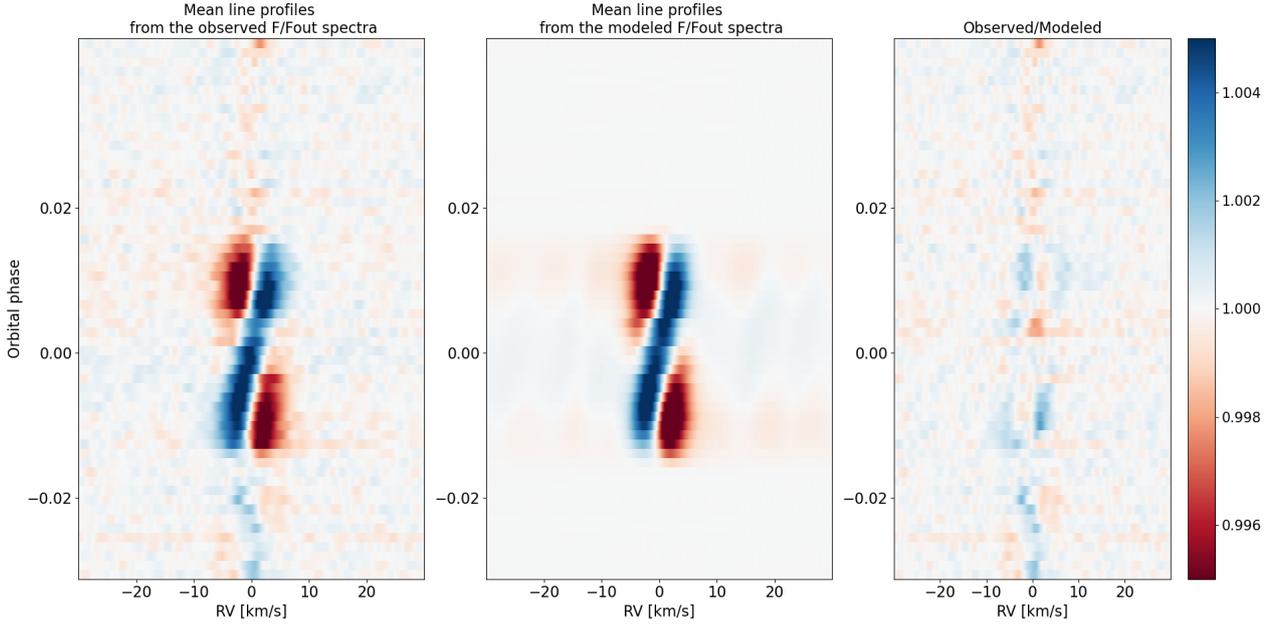

**Figure 5.** The mean line profile of 262 stellar lines for single transmission spectra derived from the observations (left) and the model (center) in the stellar restframe. The right panel shows the observation divided by the model.

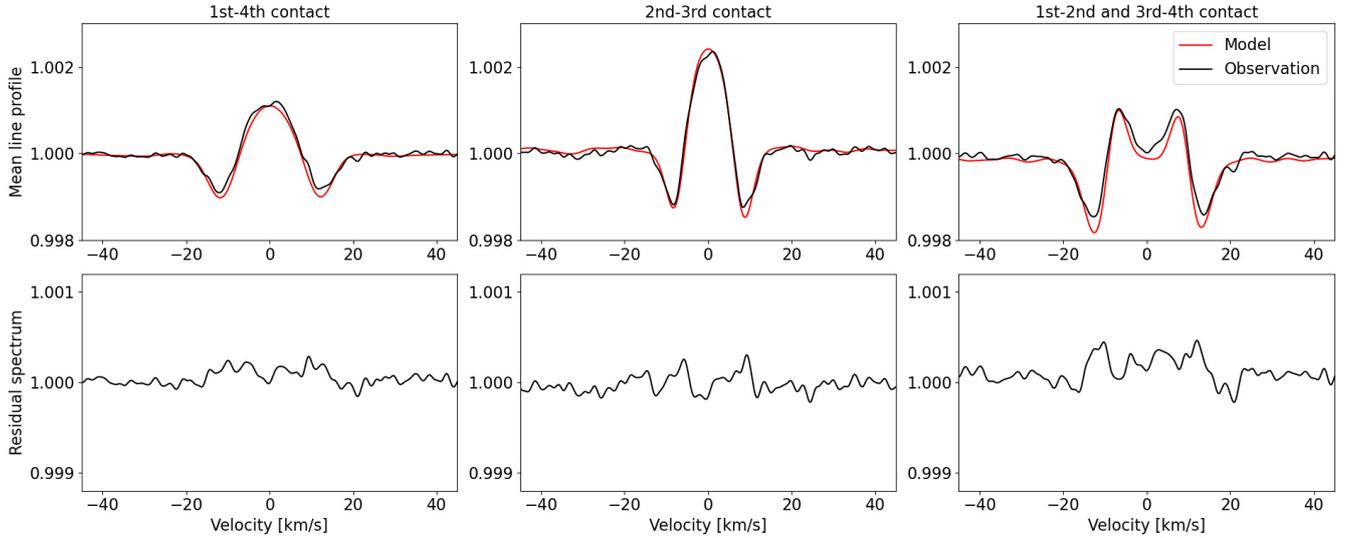

**Figure 6.** The mean line profile of 262 stellar lines for the combined transmission spectra i.e. between the 1st-4th contact (left), between the 2nd-3rd contact (center) and 1st-2nd + 3rd-4th contact (right). Top panel: Mean line profiles for the observations (black) as well as the model (red). Bottom: The residual spectra dividing the observed mean line profile by the modeled mean line profile.

### 3.1 The comparison of the modeled and observed transmission spectra

We compare the modeled and observed transmission spectra which contain the RM-CLV-signatures to verify the suitability of the applied RM-CLV-model. Figure 5 compares the mean line profile for each transmission spectrum within a 2D color map for the observed spectra (left), the modeled spectra using the LTE models (center), and their ratio (right). The observed and modeled mean line profiles show a good spectral and temporal match of the RM-CLV-signatures in the transmission spectra, not demonstrating significant scatter in the right panel. We derive in Figure 6 in addition the mean line profiles for transmission spectra (top) combined between the 1st - 4th contact (left), between the 2nd - 3rd contact (center) and 1st-2nd + 3rd-4th contact (right). For the modeling, we used a $v \sin i$ = 2.8 km/s which showed a better match between the models and the observations derived by comparing the modeled and observed mean line profiles via a $\chi^2$-analysis for different $v \sin i$ values. The bottom panels show the ratio between observation and model. The modeled and observed mean line profiles in the combined transmission spectra demonstrate also a good match for the observed and modeled RM-CLV-feature, with a slight scatter.

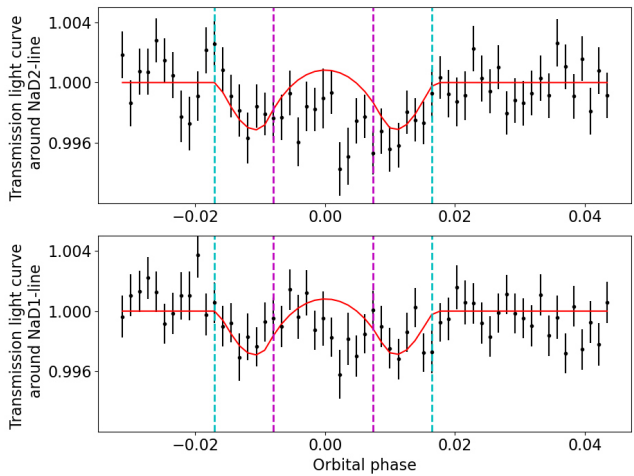

**Figure 7.** Transmission light curve showing integrated flux at a bandwidth of ± 0.4 Å around the Na D-lines. The solid red line shows RM-CLV model. Cyan dashed lines mark the 1st - 4th contact and the purple dashed lines mark the 2nd - 3rd contact. Top: Light curve at the Na D2- line. Bottom: Light curve at the Na D1- line.

## 4 ANALYSIS AND RESULTS: THE NA D-ABSORPTION

### 4.1 The transmission light curves around the Na D- lines

We derive the transmission light-curve around the Na D- lines by integrating the normalized flux in each spectrum at the stellar Na D-cores within a bandwidth of ± 0.4 Å, thus approximately ± 20 km/s (the region around the expected planetary Na D- absorption) and divide this by the mean flux value of the out-of-transit spectra. The error bars are calculated via uncertainty propagation and scaled with respect to the standard deviation of the out-of-transit data to avoid underestimation.

Figure 7 shows the transmission light curves around the stellar Na D2-line (top) and Na D1-line (bottom). The dashed lines show the contact points 1st – 4th (cyan) and 2nd – 3rd (purple). Some Na D absorption is visible for both lines around mid-transit phases between the 2nd - 3rd contact. The absorption is absent during the ingress and egress phases, similar to what is presented in Borsa & Zannoni (2018) (see their Figure 8). At these positions, the RM-CLV-model matches mainly with the observation. We investigate in the upcoming sections further the Na D- absorption properties by deriving the transmission spectra.

### 4.2 The transmission spectra at different orbital phases

We create the transmission spectra dividing the in-transit spectra by the mean out-of-transit spectrum to infer the Na D-absorption. Due to the planetary orbital motion, a Doppler-shift is introduced to any atmospheric feature with respect to the orbital phase. We search within the transmission spectra for the Doppler-shifted Na D absorption at different orbital phase positions within a 2D color map (see e.g. Yan et al. 2019).

Figure 8 shows the 2D color map of the transmission spectra -1 divided by the out-of-transit column standard deviation at each pixel for the Na D2-line (top) and Na D1-line (bottom), similarly to the approach presented in Birkby et al. (2013) and Casasayas-Barris et al. (2021). The column out-of-transit standard deviation is a measure of the scatter in the spectrum, obtained as the standard deviation of pixels at each wavelength bin. Due to the significant difference in the S/N ratio values at the Na D-line cores and wings, this step enables the illustration of the 2D maps with better scaling for the statistical scatter, i.e. noise normalized. The left panel shows the observed transmission spectra not corrected from the RM-CLV-model and the right panel shows the observed transmission spectra divided by the modeled RM-CLV- transmission spectra. Inspecting the color maps before correcting by the RM-CLV-model (left panels), unambiguous detection of absorption of planetary origin is not possible. Following the expected trajectory of planetary absorption (black dashed line), a slightly stronger decrease in flux hinting on absorption becomes visible at the line core region at 0 km/s after correcting with the RM-CLV-model for both Na D-lines. Due to the low planetary Doppler-velocity of around ∓ 7 km/s at phases between the 2nd - 3rd contact, it is challenging to distinguish if the absorption is of stellar or planetary origin. Especially the absence of significant absorption at the ingress and egress phases raises doubts on the planetary origin, where the RM-CLV-effect plays a less significant role as demonstrated in Figure 4. The absence might be due to the weaker absorption imprint of the partially occulted planetary atmosphere during the ingress and egress phases in the transmission spectra. We derive therefore combined transmission spectra to investigate further on the Na D-absorption.

### 4.3 The combined transmission spectrum

To resolve the planetary Na D-absorption, each in-transit transmission spectrum is shifted according to the expected planetary radial velocity back to the same restframe and combined. The ingress and egress phase of the HD 189733b transit last for around 50% of the full transit duration. To show the absorption feature at different orbital phase positions, we derive *the combined transmission spectrum for observations between the 1st-4th contact* (which we denote as $T_{1st-4th}$) combining 27 spectra, *the combined transmission spectrum for observations between the 2nd-3rd contact* (i.e. the spectra without the ingress and egress phases which we denote as $T_{2nd-3rd}$) combining 13 spectra as well as *the combined transmission spectrum for observations between the 1st-2nd and 3rd-4th contact* (i.e. accounting for the spectra acquired only during the ingress and egress which we denote as $T_{1st-2nd+3rd-4th}$) combining 14 spectra.

Figure 9 shows (top) $T_{1st-4th}$ (left), $T_{2nd-3rd}$ (center) and $T_{1st-2nd+3rd-4th}$ (right) not corrected from the RM-CLV-model and (bottom) divided by the RM-CLV-model. The top panels show a decrease in the flux level at the Na D-line positions for $T_{1st-4th}$ and $T_{2nd-3rd}$ not following the RM-CLV-model. This is stronger for $T_{2nd-3rd}$ compared to $T_{1st-4th}$, whereas the flux level in $T_{1st-2nd+3rd-4th}$ matches the RM-CLV-model quite well, indicating again the absence of absorption at the ingress and egress phase positions in our data. The RM-CLV-profiles show much larger amplitude for $T_{2nd-3rd}$ compared to $T_{1st-4th}$. The bottom panels shows the observation corrected from the RM-CLV-effect. The absorption lines differ strongly from each other, being absent at $T_{1st-2nd+3rd-4th}$, and shallower in $T_{1st-4th}$ compared to $T_{2nd-3rd}$ most probably due to the absence of absorption at the ingress and egress positions. Moreover, the amplitude of the modeled RM-CLV-profile mainly determines the amplitude of the feature. The $T_{2nd-3rd}$ shows a feature around the expected Na D line positions, hinting at planetary absorption. The FWHM of the Na D1- feature is significantly smaller than the line width one would expect accounting for natural line broadening and broadening by the planetary rotation, making the absorption dubious. Table 1 shows the derived line shape properties for the Na D-

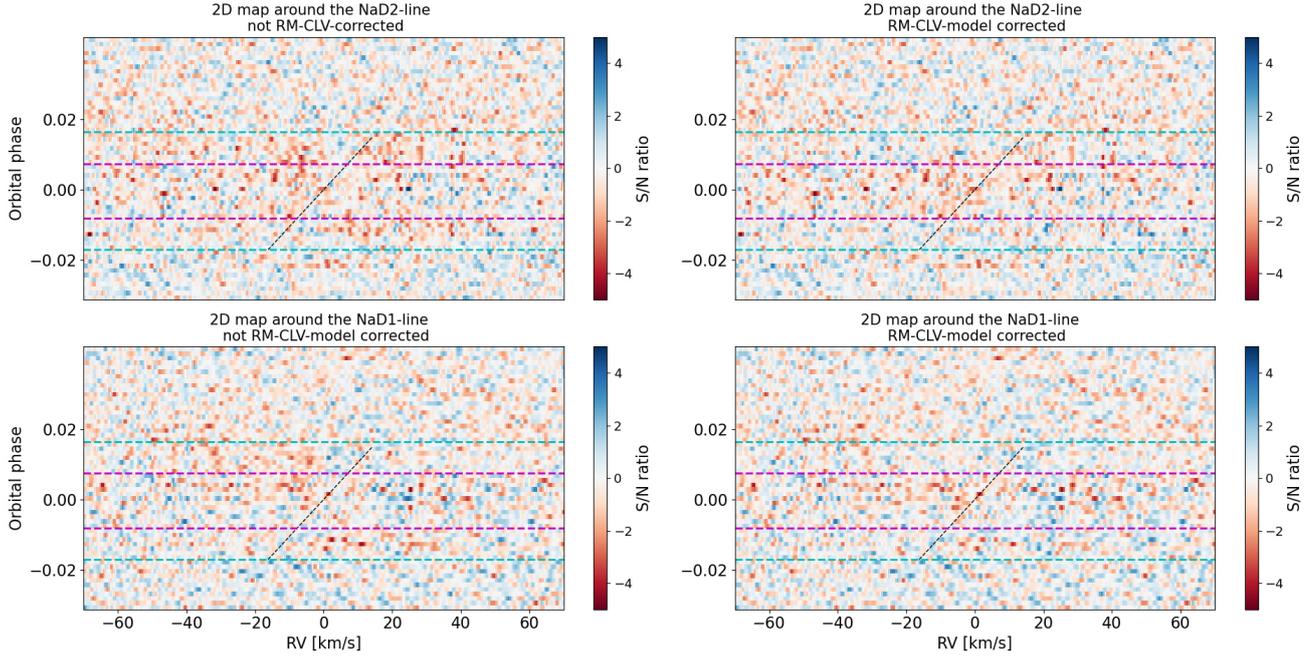

**Figure 8.** 2D color map showing the transmission spectra - 1 divided by the out-of-transit column standard deviation at each pixel at different orbital phases for the Na D2 line (top) and Na D1 line (bottom) in the stellar restframe. Cyan dashed lines mark the 1st - 4th contact and the purple dashed lines mark the 2nd - 3rd contact. The black dashed lines illustrate the planetary absorption trajectory. Left: Transmission spectra before correcting with RM-CLV-model. Right: Transmission spectra corrected with the RM-CLV-model.

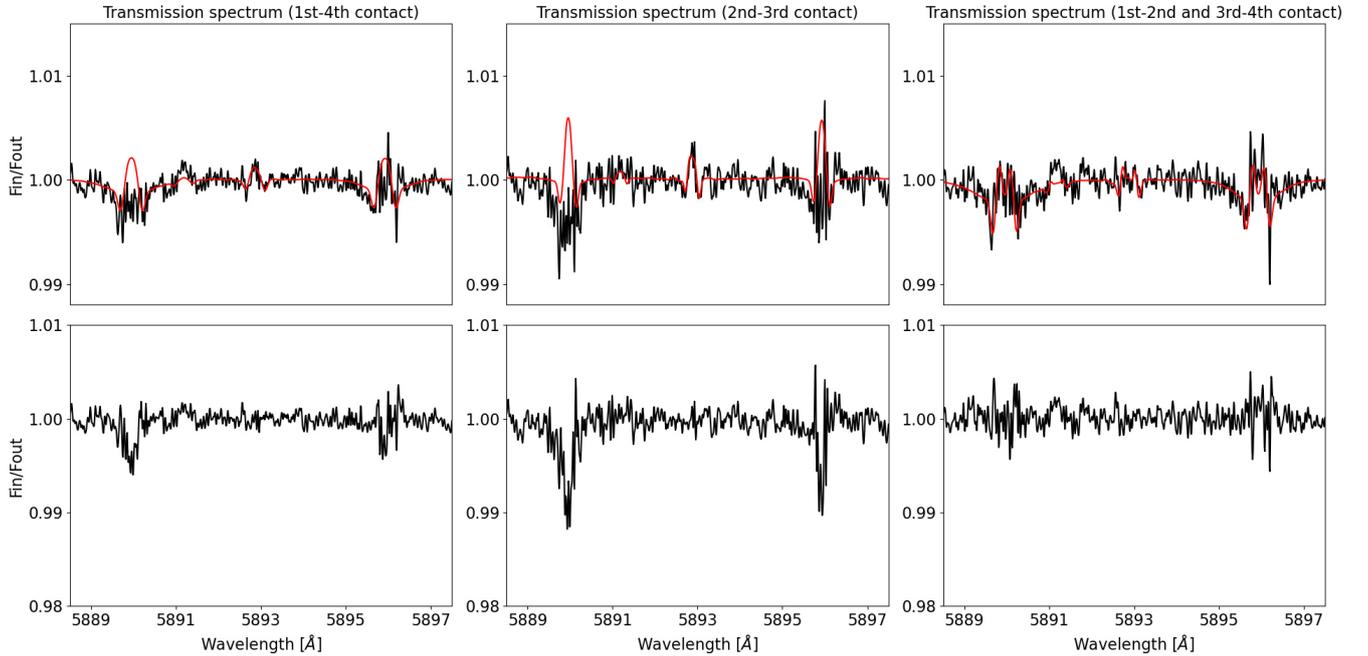

**Figure 9.** Combined transmission spectrum in the planetary restframe around the Na D-lines for $T_{1st-4th}$ (left), $T_{2nd-3rd}$ (center) and $T_{1st-2nd+3rd-4th}$ (right). Top: Combined transmission spectra (black) before correcting from the RM-CLV-model (red). Bottom: Combined transmission spectra after correcting for the RM-CLV-model.

**Table 1.** Na-D line properties from the HARPS datasets from different works and PEPSI data investigated in this work

| Work | Wyttenbach et al. (2015) | Casasayas-Barris et al. (2017) | Langeveld et al. (2021) | Sicilia et al. (2022) | This work |
|---|---|---|---|---|---|
| Na-D2 line | | | | | |
| FWHM [Å] | 0.52 ± 0.08 | 0.64 ± 0.04 | 0.46 ± 0.04 | 0.49 ± 0.08 | 0.28 ± 0.02 |
| LC [%] | 0.64 ± 0.07 | 0.72 ± 0.05 | 0.64 ± 0.07 | 0.49 ± 0.09 | 0.96 ± 0.05 |
| $v_{shift}$ [km/s] | | | -4.13 ± 1.53 | | -0.72 ± 0.41 |
| | | | | | |
| Na-D1 line | | | | | |
| FWHM [Å] | 0.52 ± 0.08 | 0.60 ± 0.06 | 0.41 ± 0.05 | 0.54 ± 0.09 | 0.13 ± 0.02 |
| LC [%] | 0.40 ± 0.07 | 0.51 ± 0.05 | 0.53 ± 0.07 | 0.43 ± 0.08 | 0.88 ± 0.12 |
| $v_{shift}$ [km/s] | | | 0.31 ± 2.05 | | -0.76 ± 0.47 |
| | | | | | |
| $v_{mean}$ [km/s] | -8 ± 2 | -2 | -1.8 ± 1.2 | -1.9 | -0.74 ± 0.45 |

Note: The values from this work are shown for $T_{2nd-3rd}$. The comparison with Langeveld et al. (2021) is done with their values for the telluric correction applied with Molecfit (see their Table 2), which led to deeper absorption lines. The velocity shift values have been calculated according to their reported shift values and the central wavelengths at 5889.951 Å and 5895.924 Å. The mean velocity shift is taken from their work. The values from Sicilia et al. (2022) are taken from their Table 2 and consider CLV- and RM- correction and are converted from km/s into Å with mean error values. Casasayas-Barris et al. (2017) did not provide an uncertainty on the line shift value.

features derived from the $T_{2nd-3rd}$ with a more detailed discussion on the line shape properties provided in Section 5.1. The absence of major absorption during ingress and egress is in agreement with the absence of absorption in the transmission light curve (Figure 7) as well as in the 2D color map (Figure 8). The explanation for this remains an open question, hinting at strange dynamics in the planetary atmosphere of HD 189733b if the absorption is of planetary origin.

### 4.4 Investigating the origin of the Na D-absorption

We investigate the origin of the Na D- absorption feature by investigating the variability of first the stellar Na D- lines and second the planetary Na D- absorption feature.

We investigate if variability is arising at the low S/N ratio region around the chromospheric stellar Na D line cores which might affect the planetary Na D- feature. For instance, stellar activity can introduce variability to chromospherically sensitive lines and thus can affect the transmission spectra (Cauley et al. 2018). Such variability hints at stellar spurious effects as we do not expect variability in absorption arising in the planetary atmosphere within a short timescale of a transit. To study such spurious features, we mask the line core region in the transmission spectra around the stellar Na D-lines in the stellar restframe in steps of ±2 km/s and recalculate the combined transmission spectra and the RM-CLV-models in the planetary restframe. Figure 10 shows $T_{1st-4th}$ (first column), $T_{2nd-3rd}$ (second column) and $T_{1st-2nd+3rd-4th}$ (third column). The fourth column illustrates the masked region around the stellar line cores. To show the difference between the profiles arising at $T_{2nd-3rd}$ compared to $T_{1st-4th}$ as well as $T_{1st-2nd+3rd-4th}$, we additionally show the RM-CLV-profile combined with the planetary Na D- absorption (dashed lines) derived for $T_{2nd-3rd}$ (applying a Gaussian fit to the Na D- features from the central panel in Figure 9, see Section 5.1 for further details). The planetary absorption is introduced in the RM-CLV-model to account for the partial occultation by the atmospheric annulus at ingress/egress to avoid overestimation of the expected absorption signature. For all cases, one can see that the impact of the RM-CLV-effect (solid lines) becomes smaller for masking the stellar line core regions. For $T_{1st-4th}$ and $T_{1st-2nd+3rd-4th}$, our data do not follow the RM-CLV × absorption profile, showing that the planetary Na D absorption must be different compared to $T_{2nd-3rd}$. As the RM-CLV- model without planetary absorption matches the observation for $T_{1st-2nd+3rd-4th}$ for the different core masking scenarios, variability at the Na D- line core regions at these orbital phase positions is unlikely. Moreover, if absorption had been present during the ingress and egress, this should have manifested itself in the spectra (dashed lines), which we do not see in our analysis. For $T_{2nd-3rd}$, the observed spectra do not follow any of the models. This hints at the variability of the stellar Na D- lines. However, if the variability had arisen from chromospheric activity manifesting itself in the line core region, the data would probably follow the RM-CLV-profile without planetary absorption for the different core masking scenarios. The origin of the variability thus is unlikely to arise from this, hinting at a different unknown source.

We further investigate the origin of the Na D- feature by investigating the variability of the planetary absorption feature. Figure 11 shows $T_{2nd-3rd}$ for different planet velocity semi-amplitude Kp-values with the RM-CLV × absorption profile (dashed lines). Due to the low planetary Doppler-velocities at the in-transit phase positions, contribution from the planetary Na D- absorption even for these different Kp-values is still expected. However, the planetary absorption should be non-variable within the short timescale during the 2nd-3rd contact and therefore the observation and model should match in the combined transmission spectra computed for different Kp-values, with the absorption profile smearing out and becoming shallower. The first row shows $T_{2nd-3rd}$ for the expected planetary Kp-value, the second row shows $T_{2nd-3rd}$ for 0×Kp i.e. the stellar restframe, the third row shows the same for -1×Kp and the fourth row shows the case for -2×Kp. The first column shows the result around the Na D2-line, the second column shows the result around the Na D1-line and the third column shows the observation divided by the RM-CLV × absorption model. For the different Kp- values, although the model and data seem to match, a strong scatter is still visible in the residual spectra, especially in the second and third row at the Na D2- feature, prohibiting an unambiguous verification of the planetary origin of the sodium feature. The absorption seems to be even stronger in some cases compared to the expected Kp-value, but in the case of constant absorption, the scatter should be at the same level. This hints again at variable absorption if it's originating from the planetary atmosphere. However, due to the strong overlap between the stellar surface radial velocities and the planetary Doppler-velocities for $T_{2nd-3rd}$, note that any spurious absorption with stellar origin arising at the stellar rest-frame could mimic a similar absorption profile, making more observations required to make a clear statement on the origin of the Na D- feature from our dataset.

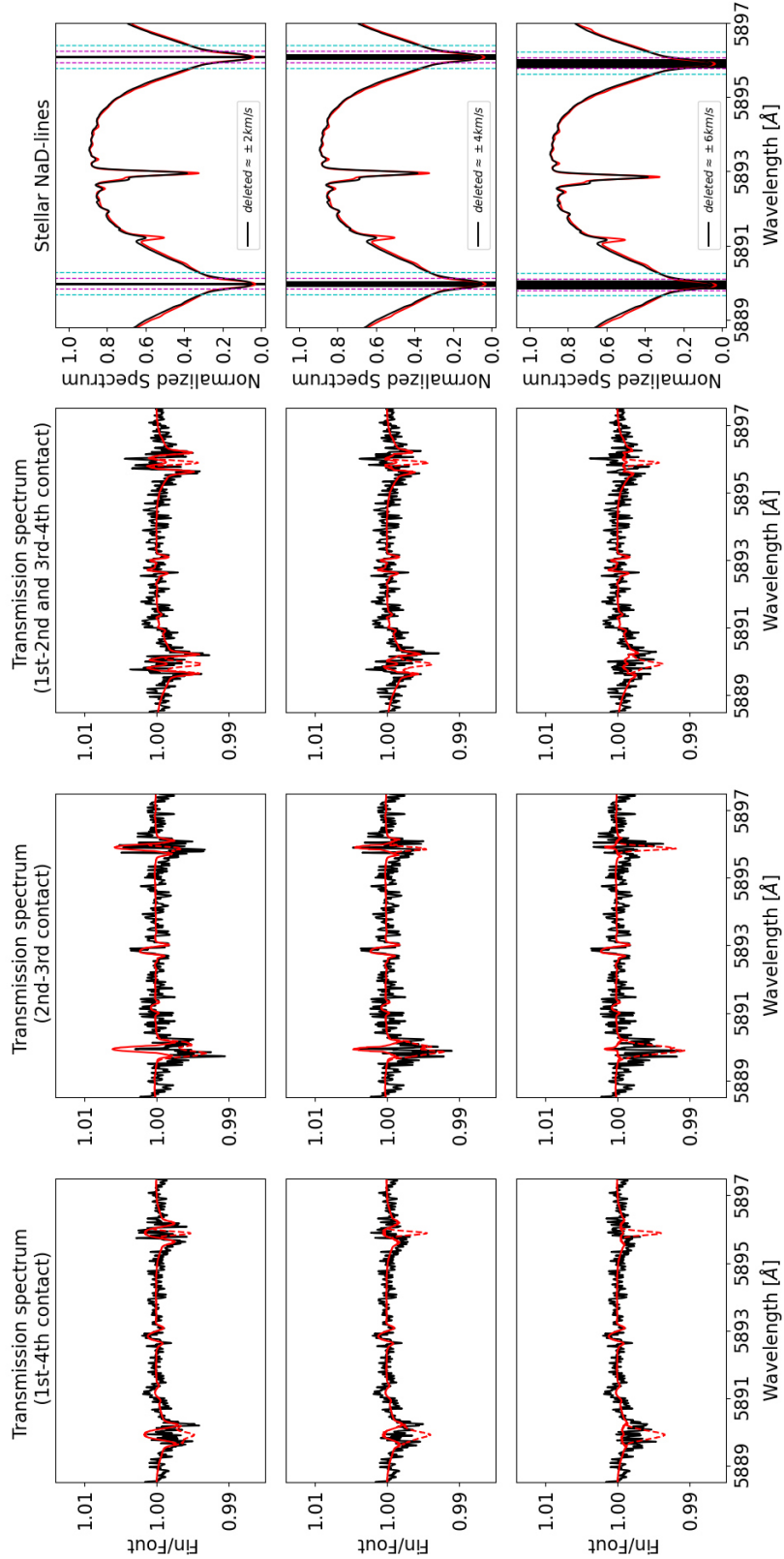

**Figure 10.** Investigating the origin of Na D-absorption on HD 189733b. Combined transmission spectrum in the planetary restframe around the Na D-lines for $T_{1st-4th}$ (first column), $T_{2nd-3rd}$ (second column) and $T_{1st-2nd+3rd-4th}$ (third column). The solid red line shows the RM-CLV model. The dashed line shows the RM-CLV × absorption model including the Na D- absorption derived from $T_{2nd-3rd}$ (see Table 1). The fourth column shows the stellar Na D- lines and the masked region in the stellar restframe. The different rows show the effect of masking the line core regions, i.e. ± 2km/s (top), ± 4km/s (center), and ± 6km/s (bottom). Cyan dashed lines mark the 1st - 4th contact and the purple dashed lines mark the 2nd - 3rd contact to demonstrate the correlation between the masked region and the region of planetary Na D- absorption.

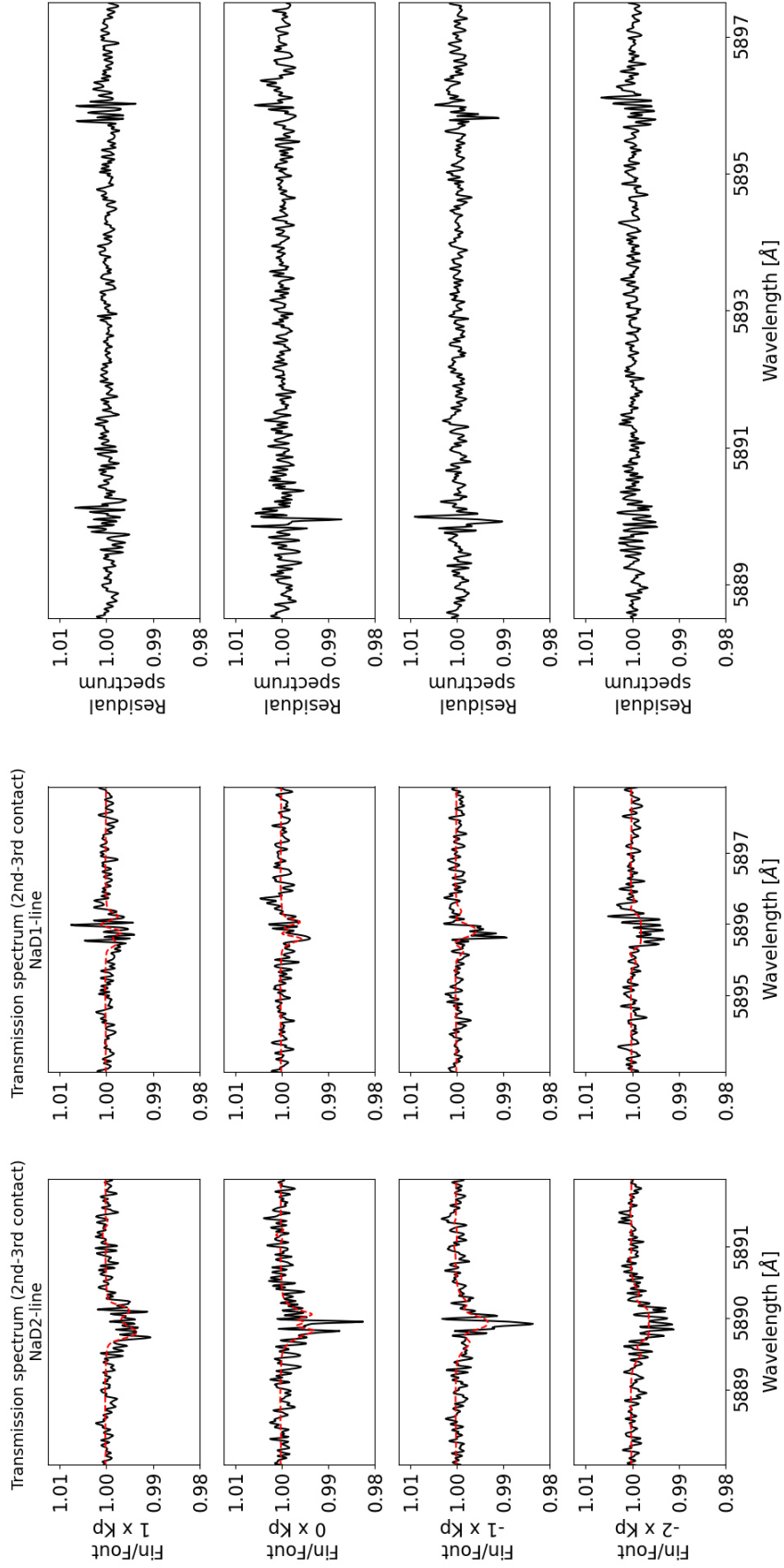

**Figure 11.** Combined transmission spectrum in the planetary restframe for $T_{\text{2nd-3rd}}$ around the Na D2-line (left) and Na D1-line (center). The red dashed line shows the RM-CLV × absorption model. The right panel shows the ratio of the observed and computed transmission spectra. Each row refers to different Kp-values.

## 5 DISCUSSION

### 5.1 The line shape parameters

In the following, we go on the assumption that the observed absorption in the $T_{2nd-3rd}$ is planetary in origin. We derive the properties of the Na D- feature from $T_{2nd-3rd}$ applying a Gaussian fit. The line shape properties derived in this work for the Na D2-line after applying RM-CLV-correction show a FWHM = 0.28 ± 0.02 Å, LC = 0.96 ± 0.05 % and $v_{shift}$ = -0.72 ± 0.41 km/s. The line shape properties for the Na D1-line after applying RM-CLV-correction show a FWHM = 0.13 ± 0.02 Å, LC = 0.88 ± 0.12 % and $v_{shift}$ = -0.76 ± 0.47 km/s. Figure 12 shows for comparison the LC (top), the $v_{shift}$ (center), and the line FWHM (bottom) for the Na D2- line (left) and the Na D1-line (right) derived in this work (green shaded region) and previous HARPS investigations (blue shaded region).

The bottom panels show in addition the wind speeds required to broaden the planetary absorption lines (dashed lines) calculated using a 100 × 100 grid model, mapping each pixel on the planet's surface (see Keles et al. 2019, 2020) for further information). The grid model contains in each pixel the unbroadened Gaussian Na D-absorption line profile with LC = 1.1% and FWHM = 0.18Å, which we derived using the Na D2-line properties which correspond to an atmosphere with no streaming global wind but the planetary rotation. The LC corresponds to an atmospheric extension of 1.00 × $R_{pl}$ to 1.22 × $R_{pl}$. We put on top of the planetary rotation velocity the global sr with different wind velocities. We Doppler-shift the profiles in each pixel with respect to the planetary rotation velocity ($v_{rot}$) and the global sr wind velocity ($v_{sr}$) streaming in the same direction (see the caption of Figure 12 for further information). We derive the broadened line profile from the grid model and determine its FWHM to identify the broadening effect for different wind strengths. Although the derived values are first-order approximations that require 3D modeling for more accurate estimations (Kempton et al. 2014), note that a global streaming sr wind is an optimistic scenario requiring lower wind speeds to introduce similar broadening to the absorption lines compared to e.g. zonal restricted winds, winds flowing opposite to the planetary rotation, winds showing non-uniform velocity distribution or vertical upward winds.

Investigating the Na D2-absorption for $T_{2nd-3rd}$ (left panel), the derived line shift suggests a dn wind velocity of around 0.7 km/s and the derived FWHM suggests a sr wind velocity in the order of 3-4 km/s on top of the planetary rotation on HD 189733b.

Inspecting the Na D1-absorption for $T_{2nd-3rd}$ (right panel), similar LC and dn wind velocities are inferred. Significant difference arises with respect to the inferred FWHM of the Na D1- line, which is much narrower than the FWHM derived for the Na D2-line, and even in comparison to what would be expected only accounting for temperature- and pressure broadening. This raises again doubts of the planetary origin of the Na D- absorption.

Overall, the results inferred with respect to the planetary winds should be considered cautiously, as they rely on the assumption that the presented Na D- absorption is of purely planetary origin as well as not affected by stellar activity or other nuisance effects, which is of question.

### 5.2 Comparison to other studies

The Na D-line absorption on HD 189733b has been analyzed using the three HARPS transit observations by several studies in which significant Na D absorption around both resonant Na D- lines have been inferred. Figure 12 (blue shaded region) and Table 1 shows the LC, FWHM and $v_{shift}$ of the Na D lines inferred by a Gaussian fit from works which have provided all values in their manuscripts, i.e. from Wyttenbach et al. (2015), Casasayas-Barris et al. (2017), Langeveld et al. (2021) and Sicilia et al. (2022). As the line properties inferred in the HARPS data are derived from combined transmission spectra including the ingress and egress phases, we show the line shape properties from the PEPSI spectra (see Figure 9) derived for $T_{1st-4th}$ in addition to the line shape properties derived for $T_{2nd-3rd}$. Including spectra from the ingress and egress phases, the PEPSI data and HARPS investigations show similar values for the LC, agreeing mostly with their uncertainties, whereas the LC inferred for $T_{2nd-3rd}$ in the PEPSI data is significantly stronger. Overall, the HARPS data investigations show a tendency of higher wind velocities, for dn winds and winds which introduce line broadening, compared to the observations with PEPSI, although the dn wind agrees within 1$\sigma$ with the mean value derived by Sicilia et al. (2022) and Langeveld et al. (2021).

The main difference between the PEPSI and HARPS data arises for the derived FWHM of the Na D- lines. If the absorption is of planetary origin, with respect to strong sr or vertical upward winds, the new PEPSI data suggests a breezy planet with weaker winds, whereas the older HARPS data suggest a stormy planet with stronger winds. Figure A1 shows for comparison the PEPSI and HARPS nights showing the S/N ratio per combined pixel (top) and the RM-curve (bottom). Cyan dashed lines mark the 1st - 4th contact and the purple dashed lines mark the 2nd - 3rd contact. The first HARPS night contains only a few in-transit spectra due to longer exposure time, the second HARPS night suffers from the lack of pre-transit observations and the third HARPS night has significantly higher S/N ratio values in the spectra observed during the transit compared to the out-of-transit observations. Although the difference in FWHM derived for the PEPSI and previous HARPS investigations might be due to atmospheric variability with respect to the difference in observation time which differ by more than a decade between the datasets, another possibility might be reasoned by the data quality. Although HARPS is a more stable instrument compared to PEPSI, the observing conditions have not been perfect for the HARPS nights. Another possibility for the differences might be the issue of the origin of the Na D- features, which cannot be unambiguously attributed purely to the planetary atmosphere based on the data at our disposal. Furthermore, stellar activity related issues can affect transmission spectra (Cauley et al. 2018) and the inferred results from different observations. The comparison to previous works investigating the Na D-absorption on HD 189733b shows that the inferred wind properties of this planet deviate strongly for different datasets.

Comparing the findings to studies deriving wind properties of other elements, the stellar activity of HD 189733 challenges the accurate detection of wind properties. Although the possible evidence of dn winds and sr winds have been inferred using the H$\alpha$ absorption (Cauley et al. 2017b) and the He absorption (Salz et al. 2016), stellar activity related variability challenges the derived wind properties (Cauley et al. 2016, 2017a; Salz et al. 2016). Other elements such as $H_2O$ and CO absorption (Brogi et al. 2016, 2018) show no significant presence of a sr wind, but the possibility of a dn wind, which consistent with its absence at a $1.5\sigma$ level. Similar results are presented for the K I- absorption (Keles et al. 2020), showing no significant winds on HD 189733b. If the presented Na D-absorption is purely of planetary origin, our findings agree with the latter works, demonstrating no strong winds on HD 189733b.

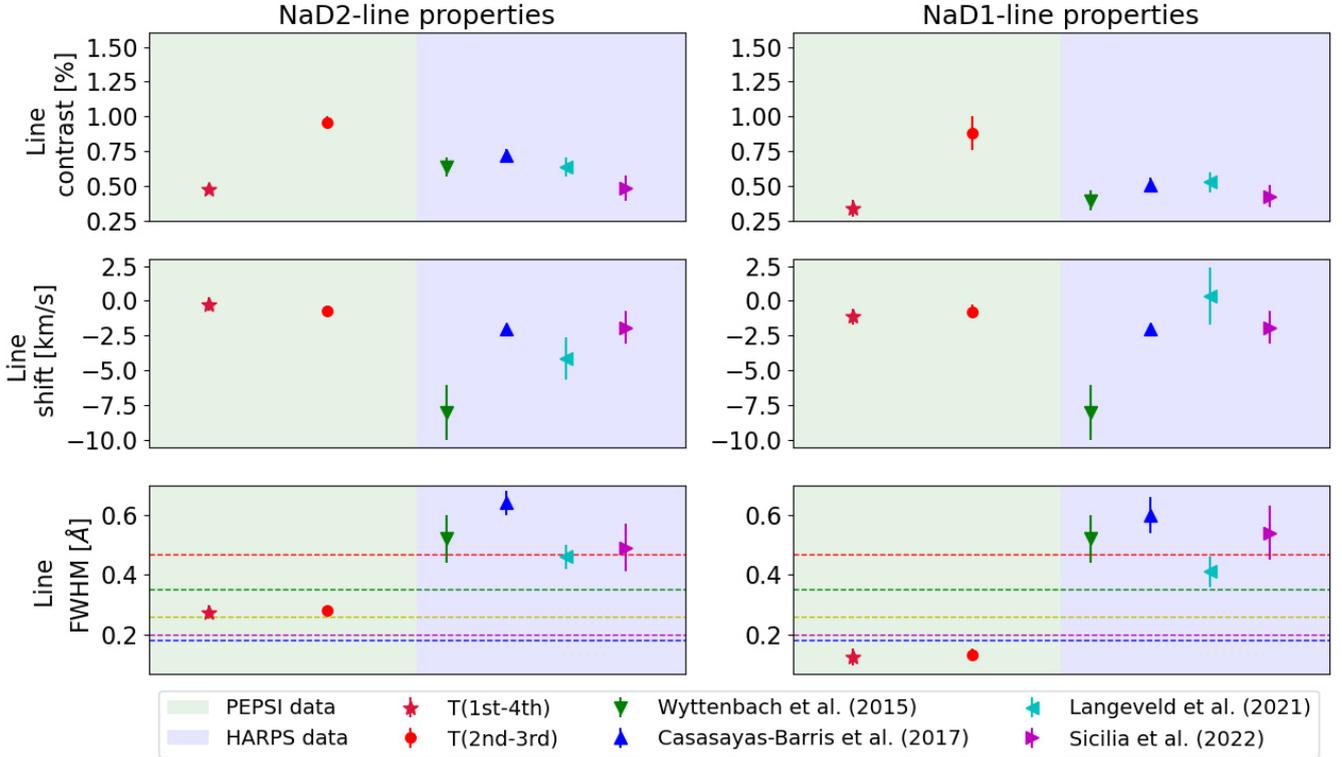

**Figure 12.** Effect of the RM-CLV-correction on the inferred Na D2 (left) and Na D1 (right) line shape properties, showing the line contrast (top), line shift (central), and line width (bottom). Dashed lines show no broadening (blue), only planetary rotation $v_{rot}$ (purple), $v_{sr}$ (3km/s) + $v_{rot}$ (yellow), $v_{sr}$ (6km/s) + $v_{rot}$ (green) and $v_{sr}$(9km/s) + $v_{rot}$ (red). Shown are the line shape properties derived in this work for $T_{1st-4th}$ and $T_{2nd-3rd}$, as well as for previous HARPS data investigations. See legend for further information.

### 5.3 Possible effects affecting line shape properties

Different effects might affect the derived line shape properties from transmission spectra, such as the stellar properties used to derive the synthetic stellar spectra. As demonstrated in Dethier & Bourrier (2023), changing the Na D-abundance can change the RM-CLV-profile. We tested the impact of this repeating the analysis using solar abundance value of A(Na) = 6.24 dex in our models. Note, that a strong mismatch between the observed and computed stellar Na D-line wings arose due to the difference in abundance, not affecting significantly the line core region. We repeated the analysis and inferred results matching within $1\sigma$ to the presented ones. This might be due to the circumstance that the planetary absorption takes mainly place part at the line core region. By decreasing the abundance, the stellar lines become shallower. This is also the case for increasing the stellar effective temperature. Thus, there is a temperature-abundance degeneracy, where higher temperature and lower abundance values (and vice versa) can yield very similar stellar line profiles. Therefore to account for this degeneracy, we adopted the stellar temperature to T = 4800K as well as T = 5200K and varied the abundance ±0.4 around the solar value until we matched the observed Na D-lines. We repeated the analysis from this work and inferred similar Na D line shape properties matching within $1\sigma$ with the presented ones derived in this work.

Overall, it is recommended to compare the synthetic and observed RM-CLV-feature, e.g. via the mean line profile, to avoid misleading results or even false-positive detection of planetary properties, whereas the uncertainty on the reliability of the models accounting for the center-to-limb variation will still remain for lines such as the Na D- lines. It has been shown recently that LTE stellar atmosphere models tend to overestimate the CLV-effect for solar-like stars (Reiners et al. 2023) and significant differences can arise even comparing 1D and 3D models accounting for LTE and NLTE effects (Lind & Amarsi 2024; Canocchi et al. 2024). The difference of the RM-CLV-LTE and NLTE- profiles has been demonstrated by Casasayas-Barris et al. (2021) for HD209458b around the Na D-lines, showing especially differences at the amplitudes of the RM-CLV-profiles. The dependence of the RM-CLV-profiles accounting for LTE and NLTE effects probably introduces another uncertainty to the absorption line properties derived after accounting for the RM-CLV-effect. In addtion, stellar atmosphere models do not include stellar chromospheres which might lead to effects that we are not aware of becoming especially important for planetary absorption on top of stellar line cores that originate in stellar chromospheres, such as for the Na D- lines. This might become especially important for HD 189733b for which the planetary absorption takes mainly place at the stellar line core region of the Na D-lines, as illustrated in Figure 10.

Finally, the methodology of deriving the combined transmission spectrum, for instance, accounting for broadband limb-darkening (Mounzer et al. 2022; Dethier & Bourrier 2023) or the employment of weights (Langeveld et al. 2022) can affect the derived line shape properties. If weights are used for the observations, the same must be used for the computation of the RM-CLV-profile as this will affect the properties of the RM-CLV-profile, especially for spectral lines with strong S/N ratio variations between the line core and line wings as it is the case for the Na D- lines. A more advanced approach of computing the RM-effect (Cegla et al. 2016) for instance using 3D models with simultaneous consideration of stellar and planetary (Dethier & Bourrier 2023) might be also a more robust approach to deriving the RM-CLV-profile.

## 6  CONCLUSIONS AND SUMMARY

The hot Jupiter HD 189733b is the first exoplanet that was found to have an atmosphere in a ground-based observation. The detected Na D-absorption has been thereafter investigated by several works mostly based on the same three HARPS datasets, showing the presence of day-to-night side winds as well as strong rotational winds. One important aspect becomes the RM-CLV-effect for this system, affecting significantly the transmission spectra.

We re-investigated the Na D- absorption with a new transit observation from the LBT & PEPSI with data quality being significantly better than the combined previous three HARPS data sets. The investigation shows two main outcomes:

First, our results suggests the absence of detectable in- and egress Na D- absorption following the planetary orbital trajectory, casting doubt on the planetary origin of the absorption signal detected between the 2nd and 3rd contact. Consequently, the average absorption signal differs significantly depending on whether the 1st-4th contact or the 2nd-3rd contact time period is considered. Also the inferred Na D1- absorption line width appears to be unphysically small, falling even short of the prediction accounting solely for temperature and pressure broadening.

Second, assuming that the Na D- absorption is of planetary origin, we infer weak winds on HD 189733b, showing day-to-nightside wind velocities of around 0.7 km/s and rotational winds with velocities in the order of 3-4 km/s, making HD 189733b a breezy planet at the probed atmospheric layers. The wind velocities would be in agreement with theoretical expectations (Showman et al. 2012), especially for the rotational winds, but in contradiction to previous results inferred from the HARPS datasets predicting much stronger winds. As the latter was acquired more than a decade ago, atmospheric variability might play a role. This all again is only valid, if the Na D- feature is indeed of planetary origin, which is doubtful.

The modeling of the signature of the RM-CLV-effect is crucial to interpret transmission spectrum (Casasayas-Barris et al. 2021; Dethier & Bourrier 2023), and in the particular case of HD 189733b it strongly overlaps in wavelength space with genuine atmospheric signals and has been found to be of comparable amplitude at least in the case of the Na D- lines. As the overlap of the expected atmospheric absorption and the RM-CLV-effect is comparably weak, however, during the in- and egress, a detection of absorption in these phases, would strongly support the case for a planetary origin. Unfortunately, this is absent in our analysis of the PEPSI data. The interpretation of the Na D signals in the transmission spectrum and the dynamical state of the atmosphere of HD 189733b remains fraught with difficulties.

## ACKNOWLEDGEMENTS

S. C. acknowledges the support of the DFG priority program SPP 1992 "Exploring the Diversity of Extrasolar Planets" (CZ 222/5-1).
K. P. acknowledges funding from the German Leibniz Community under grant P67/2018
A. S. B. acknowledges financial contribution from PRIN MUR 2022 (project No. 2022CERJ49).
E.S. acknowledges support from NASA Goddard
B.S.G. acknowledges the support by the Thomas Jefferson Chair for Discovery and Space Exploration at the Ohio State University

## DATA AVAILABILITY

The data underlying this article will be shared on reasonable request to the corresponding author.

# APPENDIX A:

This paper has been typeset from a T<sub>E</sub>X/L<sup>A</sup>T<sub>E</sub>X file prepared by the author.

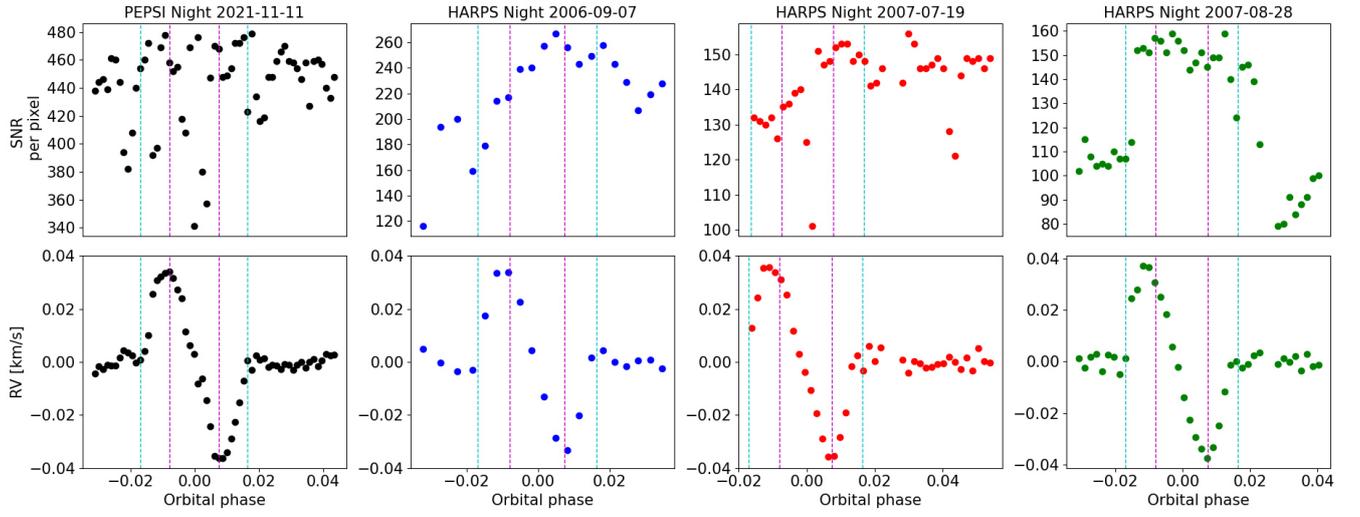

**Figure A1.** Transit observations of HD 189733b with different instruments. Top: The continuum signal-to-noise ratio value per data point in the spectra for PEPSI and HARPS data. Bottom: RM-curve. Cyan dashed lines mark the 1st - 4th contact and the purple dashed lines mark the 2nd - 3rd contact.